\documentclass[12pt]{spieman}  
\usepackage{amsmath,amsfonts,amssymb}
\usepackage{graphicx}
\usepackage{setspace}
\usepackage{tocloft}
\usepackage{lineno}
\usepackage{caption}
\usepackage{subcaption}
\usepackage{tikz}
\usetikzlibrary{arrows.meta, positioning}

\title{JWST telemetry combined with active coronagraphy: raw contrast predictions for exoplanet imaging with the Habitable Worlds Observatory}

\author[a, b*]{Raphaël Pourcelot}
\author[a]{Laurent Pueyo}
\author[c]{Emiel H. Por}
\author[a]{Marshall D. Perrin}
\author[d]{Iva Laginja}
\author[a]{Bryony Nickson}
\author[a]{Sarah Steiger}
\author[a]{Rémi Soummer}
\author[a]{Rendal Telfer}

\affil[a]{Space Telescope Science Institute, 3700 San Martin Drive, MD 21217, Baltimore, USA}
\affil[b]{Max Planck Institute for Astronomy, Königstuhl 17, Heidelberg, D-69117}
\affil[c]{Astronomy Department, University of California Santa Cruz, 1156 High St., Santa Cruz, CA 95064, USA}
\affil[d]{Universit\'e C\^ote d'Azur, Observatoire de la C\^ote d'Azur, CNRS, Laboratoire Lagrange,  Bd de l'Observatoire, CS 34229, 06304 Nice cedex 4, France}

\cftpagenumbersoff{figure}
\cftpagenumbersoff{table} 
\begin{document} 
\maketitle

\begin{abstract}
We provide a quantification of the technological gap between the James Webb Space Telescope (JWST) and the Habitable Worlds Observatory (HWO) for the goal of exo-Earth imaging around Sun-like stars at the $10^{-10}$ raw contrast level. We use JWST's in-flight telemetry of the primary segmented mirror to simulate a JWST-like telescope equipped with a modern coronagraph instrument, inspired by the Roman Space Telescope (RST) Coronagraphic Instrument (CGI), featuring an Apodized Pupil Lyot Coronagraph and active deformable mirror wavefront control on a segmented, unobstructed, off-axis telescope. We show that it can achieve $\sim10^{-10}$ raw contrast for very bright stars (brighter than magnitude 4) for fast control frequencies of 100 Hz, but that this level of control still lacks sufficient signal to correct JWST-amplitude errors for fainter stars. We show that an improvement of a factor of ten in wavefront stability is sufficient to extend this capability to a $10^{-10}$ raw contrast across all considered control frequencies (1 Hz to 100 Hz), assuming no reaction wheel vibrations, for stars up to a magnitude of 11. These results establish a new quantitative benchmark linking JWST's demonstrated thermo-mechanical stability to HWO's requirements, showing that active wavefront control relaxes the structural stability demands on the observatory, and identifying wavefront stability as the critical technological gap that must be closed for HWO to achieve its exo-Earth imaging goals.

\end{abstract}


{\noindent \footnotesize\textbf{*}Raphaël Pourcelot,  \linkable{rapourcelot@mpia.de} }


\section{Introduction}

\subsection{The contrast stability challenge associated with exoplanet imaging}

The ability to directly image exoplanets in the glare of the light of their host stars, affected by imperfections in the optics of a telescope, is a challenging problem. For existing space-based facilities this is achieved using a combination of custom masks integrated into a coronagraph instrument and image subtraction relying on the stability of datasets across a sequence of observations. For ground-based facilities this is achieved via the same coronagraphs coupled to extreme Adaptive Optics (AO) systems that correct for atmospheric turbulence, and data post-processing to calibrate residual atmospheric turbulence and drifts in the instrument. As of today, the direct imaging technique is sensitive to giant planets: historically young ones in the near infrared (IR) (up to 2 $\mu$\,m) and, more recently, mature ones in the mid-IR using the James Webb Space Telescope (JWST). At orbital scales on the order of the outer solar system, where our giant planets reside, the flux ratio between planet and star is $\sim 10^{-6}$. Breaking the giant planet barrier and accessing reflected light from smaller Earth-like planets in the Habitable Zone of nearby stars will require starlight suppression at least four orders of magnitude greater, and exquisite image stability to reach flux ratios of the order of $10^{-11}$.

One cannot polish large optics for a space telescope to the level of precision required to achieve this contrast without the help of Deformable Mirrors (DMs) to shape the static response of the instrument. In this static application, the DMs are not used in real time as they are in ground-based AO, instead they are used to set the system state at the needed level of starlight suppression using iterative algorithms. Vacuum demonstrations have shown that this can be achieved at the $10^{-10}$ level with a clear monolithic aperture \cite{Seo2019}, or $10^{-8}$ in air with a segmented aperture \cite{Soummer2024}. Segmented apertures are particularly attractive for future missions because of the low density per collecting area and flexible architecture, both of which enable fitting a larger primary mirror in a given launch vehicle. However, when compared to monoliths, segmented architectures present a much larger number of degrees of freedom for potential misalignment. Maturing technologies that provide the stability necessary for the imaging of Earth twins are one of the top priorities of the HWO technology maturation program\cite{Feinberg2026}. While the success of JWST has proven that launching a segmented space telescope is achievable, the far more stringent stability requirements for exo-Earth imaging compel a thorough reassessment of existing technologies to determine their applicability to HWO.

\subsection{Wavefront stability of JWST}
JWST high-level requirements were focused on the stability of encircled energy, and not starlight suppression. The in-flight stability, characterized with first-light data, exceeded observatory performance requirements by a substantial margin. More comprehensive analyses characterize the full temporal content available with the onboard metrology. Analysis run during commissioning\cite{Feinberg2023, Telfer2024} shows JWST's cumulative change in wavefront as a function of timescale. 

At high frequencies, of the order of $10$\,Hz, the wavefront is dominated by mechanical vibrations propagating through the backplane onto the primary segments. Those were measured during commissioning using the Fine Guider Sensor (FGS) data, obtained when the secondary mirror was out of focus in the multi-field wavefront sensing across the JWST instrument suite. That dataset is the only JWST time series fully complete at timescales longer than 7.81\,Hz, and sensitive to frequencies between $7.81$\,Hz and $40$\,Hz using prior knowledge to extract aliased information from the slower timeseries. That frequency range captures (in part) the higher harmonics of the reaction wheels, the MIRI cryocooler frequencies, and some temporary vibrations associated with spacecraft operations such as High-Gain Antenna moves. Most wheel vibrations and MIRI cryocooler contributions are below 1\,nm, with the exception of a period of time when multiple wheel components beat, which corresponds to vibrations above 5\,nm Root-Mean-Square (RMS). The High-Gain Antenna (HGA) moves can create transient segment motions that reach tens of nanometers but that settle after less than a minute. This is all in line with model predictions\cite{Schlawin2023}. 

At intermediate timescales, the JWST wavefront is dominated by the Integrated science instrument module Electronics Compartment (IEC) variations, already thoroughly discussed in \cite{Feinberg2023}. Those changes of the primary mirror state are correlated with thermal control of the attitude electronics and can be seen in wavefront sensing time series as well as in transit time series and coronagraph time series. Finally, timescales ranging from 10 minutes to hours correspond to the observatory's slower thermal responses, with combinations of segments evolving together such ad the wing pitch mode and -V3 pitch mode, specific motion modes\cite{McElwain2023, Feinberg2024} and changes in attitude of the observatory. 

All of these behaviors were predicted by pre-flight models and could in principle be remediated within a few engineering cycles for a future mission with tighter stability requirements. For JWST, such cycles were not necessary as JWST's wavefront stability already exceeds the mission requirements. Note that the wavefront sensing floor of a single exposure is about 100\,pm for either the global modes obtained in 25\,ms exposures during multi-instrument multi-field (MIMF) alignment, or the segment-level modes obtained with 8\,s exposures used to characterize longer timescales\cite{Feinberg2023}. These measurements demonstrate that JWST is stable beyond requirements; however, the sensing noise is two orders of magnitude above the projected stability required for exo-Earth imaging. Another key difference is that JWST relies on passive stability during science observations, with mirror wavefront measurement being performed weekly and control being an infrequent events performed only every few months. 

\subsection{Adaptive optics in space}

HWO will learn from JWST's architecture but will differ in several aspects \cite{Feinberg2026, Bolcar2026}. Unlike JWST, it will not be a cryogenic telescope, will not feature a mid-IR instrument, and will not feature a cryocooler and its associated excitation frequencies. It will also be actively thermally controlled unlike JWST’s optical telescope element which is passively cooled by the sunshield. As a result of this active thermal control, thermal modes such as the IEC or -V3 pitch will be controlled at the source. The thermal conductivity between the backplane and the optical surfaces will be minimized by design. It will leverage modern launch vehicles that can lift more mass per primary surface area, and thus it will be built from materials that do not transmit vibrations so easily. It will be designed for mechanical ultrastability, possibly without reaction wheels, using micro-thrusters instead, and use modern vibration isolation techniques.

Starting in 2018, NASA has invested in key component- and subsystem-level technologies toward ultra-stable telescopes with exoplanet imaging in mind, and ongoing work is continuing to mature the aforementioned technologies \cite{Feinberg2026}. Moreover, the coronagraph instrument will be equipped with DMs that can correct some of the wavefront instabilities on the fly and will keep the exoplanet imaging data as stable as possible. While those DMs are architected mostly to shape the static response of the instrument, they can also be used as an adaptive optics system in space, which has been their primary purpose for the past three decades in ground-based observatories. Several groups have been studying how to achieve sufficient levels of stability, and Potier et al.\cite{Potier2021} recently reported simulations of AO-inspired techniques stabilizing the vibration of a segmented telescope in the context of the LUVOIR study. Redmond et al.\cite{Redmond2022, Redmond2024} demonstrated that a dark hole could be maintained in air and vacuum testbeds. Pueyo\cite{Pueyo2022} discussed the fundamental limits associated with such correction and Pogorelyuk\cite{Pogorelyuk2021} introduced a method to compute accurately the ultimate closed-loop performance that may be achievable.

\subsection{This paper}

This paper combines the concept of AO in space with the JWST-measured wavefront time series and a series of numerical experiments to answer the question: if a future telescope in space had the same stability as JWST, then what would the contrast stability of such an observatory be if it was also equipped with a modern coronagraph instrument that used deformable mirrors (DMs) to correct for telescope drifts. The overall approach aims at using a similar framework to the one previously used on time series data of an instrument model for the LUVOIR mission concept\cite{Potier2021}. This consists of analytical modeling of the control loop and the coronagraph to efficiently explore the control loop parameter space and derive metrics of the achievable contrast. We first describe the simulation framework that we are using, including the mathematical description of the tools used. In a second step, we describe the processing of the time-series coefficients, the filtering of the outliers, and the Power Spectral Density (PSD) computations and modifications. Then we describe the exact instrument model that we use in the simulations. Finally we present the results and the discussion on the assumptions we made for the simulation. 

\section{Simulation procedure}

\subsection{Instrument concept}

\begin{figure}[ht!]
    \centering
    \includegraphics[width=\linewidth]{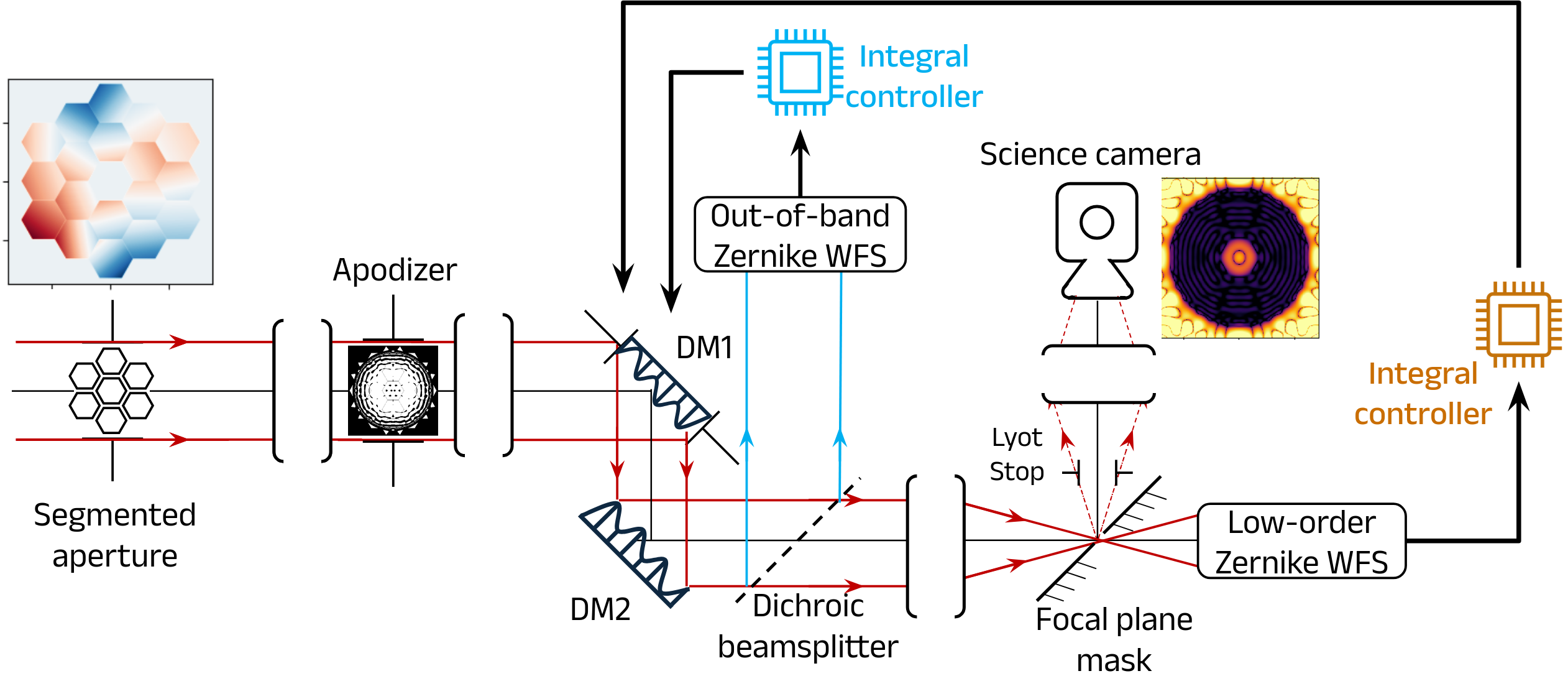}
    \caption{Schematic diagram of the simulated architecture. It includes the segmented primary mirror, an APLC coronagraph with an apodizer, a focal plane mask and a Lyot stop, two deformable mirrors, and an out-of-band and a low-order Zernike wavefront sensor. }
    \label{fig:system_schematic}
\end{figure}

The observatory and instrument overall concept studied is represented in Fig.~\ref{fig:system_schematic} and combines a large segmented aperture, for which we study the segment-level piston, tip and tilt aberrations, with a Roman Space Telescope (RST) Coronagraphic Instrument (CGI)-inspired coronagraph architecture\cite{Kasdin2020}. This includes an Apodized Pupil Lyot Coronagraph (APLC), with an apodizer optimized for the segmented aperture, a circular focal plane mask, and a circular Lyot stop. Finally, our framework also includes the ability to perform wavefront control. This implies the presence of wavefront sensors, either out-of-band using a different wavelength as the science and seeing the whole pupil including the apodizer, or as a low-order wavefront sensor\cite{Shi2016, Pourcelot2022}. Finally, to perform closed-loop control, we assume there is a phase correction device, such as a deformable mirror, that can correct for the entrance pupil segment aberrations. The parameter choices made for the simulation are summarized in Tab.~\ref{tab:simulation_param}.

\begin{table}[]
    \centering
    \begin{tabular}{|c|c|}
    \hline
    Parameter & Value \\
    \hline
    Aperture flat-to-flat diameter & D = $6.6\,m$\\
    Instrument transmission to Wavefront Sensor (WFS) & 10\,\% \\
    Central wavelength & $\lambda$ = 575\,nm \\
    Bandwidth (for photometry) & 20\,nm \\
    Number of wavelengths in the simulation & 1 \\
    Coronagraph FPM radius & $3.5\,\lambda/D$ \\
    Pure delay & 1 frame \\
    Total loop delay & 2 frames \\
    Minimum extrapolated frequency & $2\times10^{-2}$\,Hz \\
    Extrapolated PSD slope &$f^{-2}$ \\
    Low-order WFS dot diameter & $1.06\,\lambda/D$ \\
    Out-of-band WFS dot diameter & $2\,\lambda/D$ \\
    \hline
    \end{tabular}
    \caption{Main parameters used for the simulation. }
    \label{tab:simulation_param}
\end{table}

In this study, the simulations consist of finding a set of orthogonal modes of the wavefront, and then simulating a close-loop control of these modes to derive a residual variance per mode. From this residual variance, and since the number of modes is limited to the degrees of freedom of the segments of the primary aperture, it is possible to pre-compute the contrast impact of all of the modes and derive a residual contrast estimation based on the control loop parameters.

\subsection{Mathematical framework} \label{sec:math}
To estimate contrast residuals in the focal plane dark hole based on the PSD of the time series of the observatory, we use the method previously explored\cite{Potier2020}. We recall here the basic principle. 

Let $E$ be the uncorrected electric field in the image focal plane, as filtered by the coronagraph. $E$ can be decomposed into the unaberrated electric field $E_0$ plus a contribution from the primary mirror misalignments $E_{seg}$. With $N_{seg}$ segments, the wavefront can be exactly decomposed in a basis of $3\times N_{seg}$ modes, either the segment piston/tip/tilt basis or linear combinations of those. Assuming a linear regime, $E_{seg}$ can be expressed as the sum of propagation of the $3\times N_{seg}$ modes $(\phi_i)$ through the coronagraph. We call these propagated modes $(\Delta E_i)$, weighted by their $(a_i)$ coefficients in the wavefront basis.

As a result, $E$ can be written as: 
\begin{equation}
    E = E_0 + \sum_{i=0}^{3 N_{seg} - 1} a_i\Delta E_{seg, i}
\end{equation}

Consequently, with $\langle \cdot \rangle$ denoting the time average, the intensity integrated over the time series $I_{ts}$ writes:
\begin{equation}
    \begin{split}
        I_{ts} &= \langle |E|^2 \rangle \newline \\
                   &= \langle |E_0 + \sum_{i}a_i\Delta E_{seg,i}|^2 \rangle \\ 
                   &=|E_0|^2 + \sum_{i,j}\langle a_i a_j \rangle \Delta E_{seg,i} \Delta E_{seg,j}^*  + 2 \sum_i \langle a_i \rangle \mathcal{R}\{E_0^* \Delta E_{seg, i}\}\\
    \end{split}
\end{equation}

Providing the time average per coefficient is zero, one gets
\begin{equation}
    \forall i, \langle a_i \rangle = 0 .
\end{equation}

In addition, with modes that are not correlated in time, 
\begin{equation}
    \forall (i,j), i\neq j \Rightarrow \langle a_i a_j\rangle = 0 . 
\end{equation}
This absence of time correlation is ensured by the used of Singular Value Decomposition modes, that are orthogonal by construction. This computation will be detailed in Sec.~\ref{sec:svd} 
As a result, 
\begin{equation}
    \begin{split}
        I_{ts} &=|E_0|^2 + \sum_{i}\langle a_i^2\rangle \left|\Delta E_{seg, i}\right|^2
    \end{split}
\end{equation}

In the case of $(a_i)$ coefficients centered by subtracting the mean, $\langle a_i^2 \rangle$ is the variance of the coefficient over the time series. As a result, estimating residual variance after control by a closed-loop system is easily done in the Fourier domain and yields residual variance values that are immediately exploitable to estimate the integrated image over the time series. If $A_i$ is the PSD of $a_i$, then Parseval's energy conservation equality states that the variance equals the integral of the PSD: 
\begin{equation}
    \delta f \sum_{f=0}^{f_{max}} A_i(f)=\langle a_i^2\rangle,
\end{equation} 
with $f$ the frequency variable, $f_{max}$ the maximum frequency sampled by the time series, and $\delta f$ the frequency discretization step. Then, with $f_c$ the control loop frequency, $\gamma$ the loop gain, $H_{f_c, \gamma}$ the square modulus of the loop transfer function and $N_m$ the noise function introduced at the magnitude $m$, the residual variance $\langle a_i^{'2}\rangle$ after loop filtering writes
\begin{equation}
    \langle a_i^{'2}\rangle = \delta f \sum_{f=0}^{f_{max}} \left[H_{f_c, \gamma}(f) \times A_i(f) + N_m(f)\right]
 \end{equation}

\begin{figure}
     \centering
     \begin{subfigure}[b]{0.5\textwidth}
         \centering
         \includegraphics[width=\textwidth]{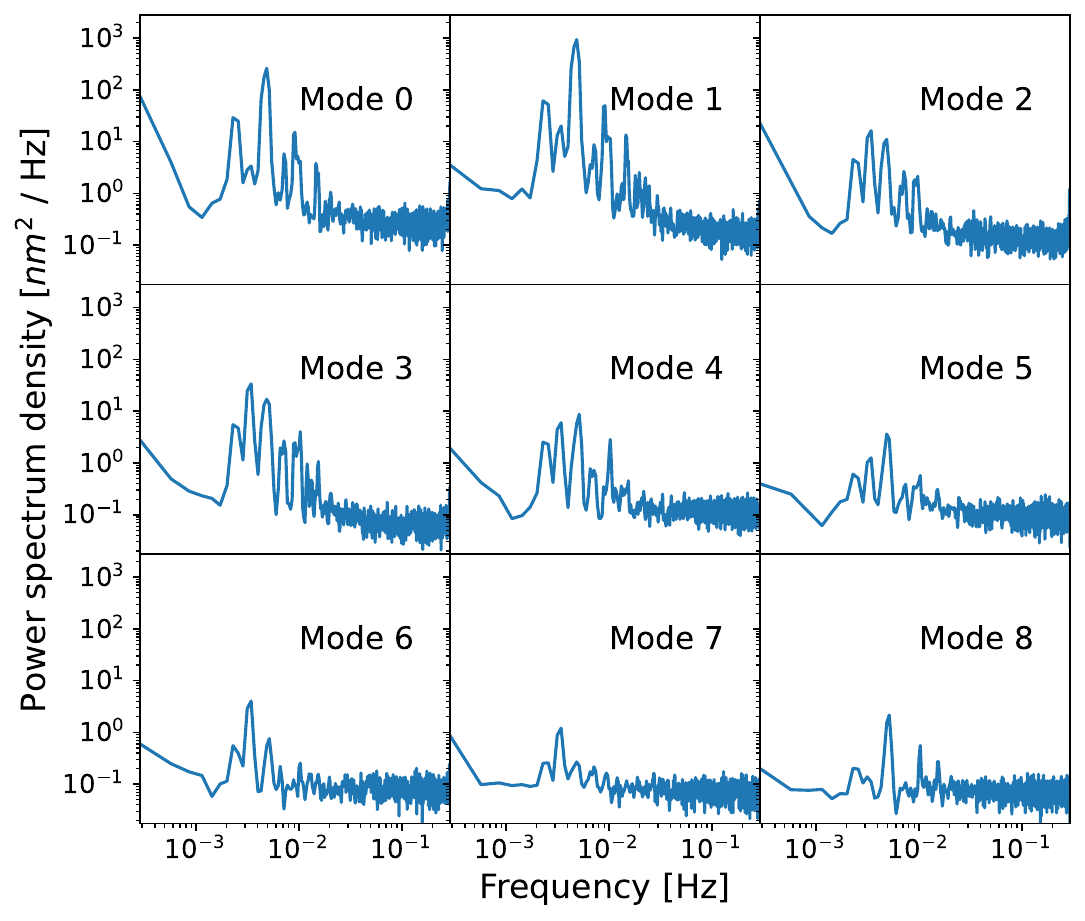}
      
     \end{subfigure}
     \begin{subfigure}[b]{0.49\textwidth}
         \centering
         \includegraphics[width=\textwidth]{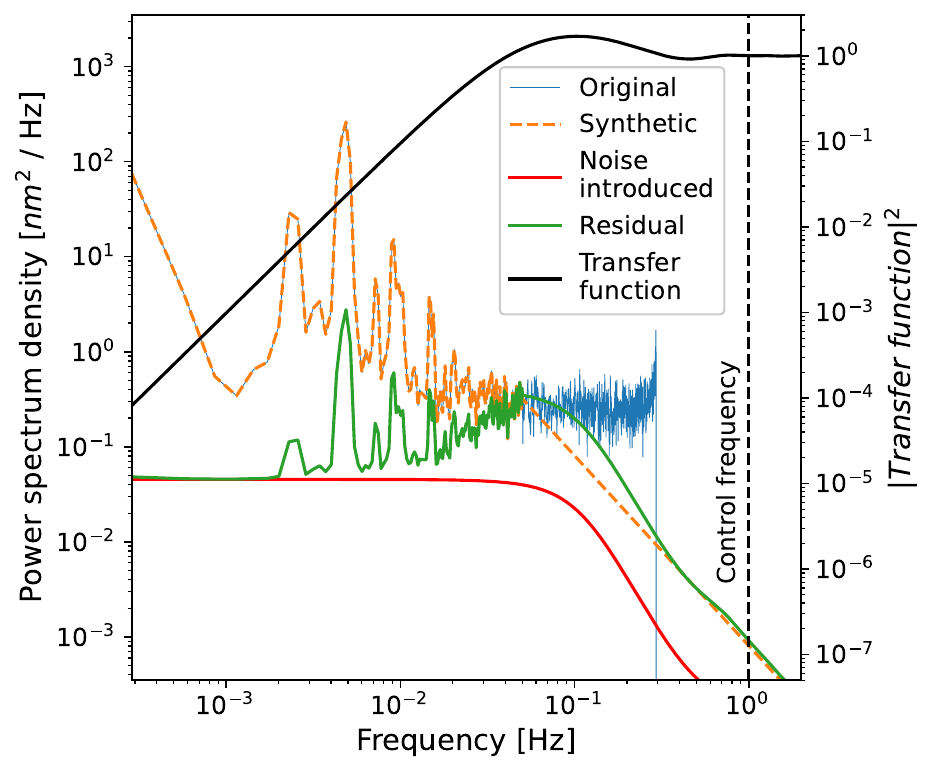}

     \end{subfigure}
        \caption{Example of PSD and control loop rejection used in the simulations. \textbf{Left}: PSD of the first 9 SVD modes as obtained by the projection of the time series onto the SVD modes. \textbf{Right}: Example processing of a modal PSD. The blue curve is the PSD of the first SVD mode. The orange curve represents the synthetic approximation and extrapolation with a power law with a -2 power. Assuming a control with a pure integrator with a gain of 0.3 running at 5\,Hz and one frame of pure delay, the black curve represents the square modulus of the transfer function. Considering a G2V star of magnitude 12, observed at 575\,nm with a 20\,nm band, the red curve represents the noise on this mode introduced by the propagation of the photon noise in the loop. Finally, the green curve represents the residual PSD filtered by the control loop, and is obtained by multiplying the synthetic PSD by the transfer function and adding the noise. }
        \label{fig:psd_world}
\end{figure}

An example of transfer functions and noise functions is presented in Fig~\ref{fig:psd_world}, where the orange, black, red, and green curve show the unaberrated PSD of a mode $A_i$, the square modulus $H$ of the transfer function, the introduced noise $N_m$ and the residual PSD $a_i^{'2}$ respectively.

Finally the residual estimated intensity $I'_{ts}$ can be calculated with 
\begin{equation}
    I'_{ts} = |E_0|^2 + \sum_i \langle a_i^{'2} \rangle \left|\Delta E_{seg, i}\right|^2 .
    \label{eq:residual_I}
\end{equation}
In this equation, we name $\Delta I$ the second part of the sum: $\Delta I = \sum_i \langle a_i^{'2} \rangle \left|\Delta E_{seg, i}\right|^2$, that is the addition of light to the unaberrated dark zone. The $\Delta I$ computation can be performed as a vector-matrix multiplication, provided the individual $(\Delta E_i)$ are stored as 1-dimensional arrays in a matrix. Finally, by defining the contrast $C$ as the intensity $I$ normalized by the peak intensity in an image acquired without the focal plane mask, we can define the contrast difference $\Delta C$ as the normalized equivalent of $\Delta I$. As such, $\Delta C$ values can be extracted by averaging $\Delta I$ over the dark zone and re-normalizing it. If this metric loses the spatial information of the speckle stability, it still provides an insightful information of the stabilization of the dark zone image. 

\subsection{Gain and frequency optimization}
The wavefront control loop is the core component of the simulation that is used to predict the stability of the observations. We model it as a closed-loop system using an integrator as the controller. As a result, the control loop behavior in the temporal frequency domain is fully described by its transfer function and noise transfer function. We consider the same block diagram and therefore the same control loop transfer functions as Potier et al.\cite{Potier2021}, but using a pure delay of the loop of only one frame, which brings the total loop delay to two frames considering the wavefront sensor and deformable mirror inherent delays. Once this is defined, the transfer function and noise transfer function are parametrized by the loop gain $\gamma$ and the control frequency $f_c$. Similarly to Potier et al., we simulate the noise $N_i$ added for each mode $i$ by assuming a white noise over the control bandpass, uniformly distributed over the wavefront sensor pixels. The noise then couples to the modes $i$ according to the covariance matrix\cite{Gendron1994a, Chambouleyron2021b}, that is the multiplication of the interaction matrix by its transpose. The uniform noise level depends on the stellar flux, computed in a 20\,nm band centered around 575\,nm to mimic RST band 1. The number of photons is calculated for a G2V star using the Exoscene package \footnote{\url{https://github.com/nasa/exoscene}} given a required magnitude, observed with an unobstructed telescope the size of JWST.

\section{JWST in-flight telemetry} \label{sec:preprocessing}

\subsection{Data package description}

The time series used in this study comes from a dataset obtained from the James Webb Space Telescope’s (JWST) Near-Infrared Camera (NIRCam) weak lens time-series observations, as part of the program 1445 (PI E. Smith), designed to characterize the optical stability of the telescope’s primary mirror assembly. The dataset consists of high-cadence imaging sequences of bright, isolated stars observed through specialized weak lens elements, WLP4 and WLP8, which introduce controlled defocus to enhance sensitivity to small wavefront errors (WFE). By deliberately blurring the point-spread function (PSF), these observations enable precise detection of segment-level misalignments, thermal drifts, and transient tilt events that would otherwise be challenging to measure in focused images.

The data were acquired using narrow- and medium-band filters (2.12\,µm for WLP4 and F212N / F210M for WLP8) with sampling intervals ranging from 1.0 to 80.8 seconds over multi-hour durations. Subarray configurations were optimized to capture the defocused PSF while maintaining sufficient signal-to-noise ratio for robust phase retrieval. Subsequent processing involved the ASPRIN phase retrieval algorithm, which decomposes the wavefront into Zernike polynomials and segment-level piston/tip/tilt terms. Known vibrational modes and systematic trends were isolated and removed, allowing for detailed analysis of both gradual thermal drifts and abrupt alignment changes.

This dataset provides a unique window into the dynamic behavior of JWST’s optical system, offering insights into the interplay between thermal effects, mechanical stresses, and vibrational disturbances. The following sections describe the observational setup, data reduction pipeline, and key characteristics of the processed wavefront measurements used in this analysis.

\begin{figure}[ht!]
    \centering
    \includegraphics[width=\linewidth]{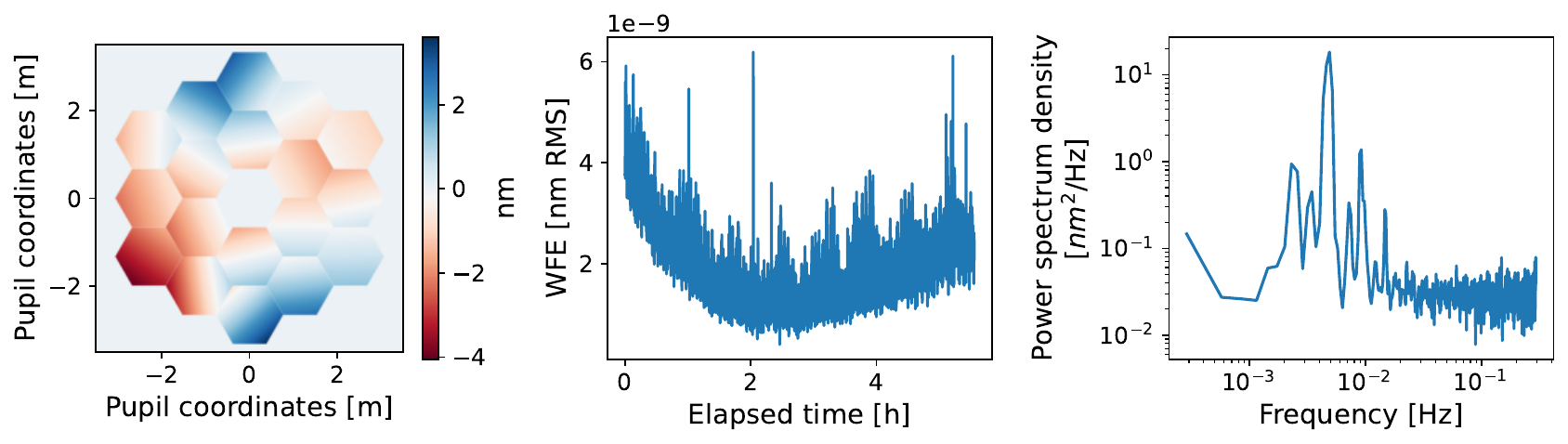}
    \caption{Example of the dynamic in-flight aberrations measured on JWST primary mirror. \textbf{Left}: Reconstruction of the primary mode measured in the time series, in surface error. \textbf{Middle}: Temporal evolution of the standard deviation on the pupil (raw data). \textbf{Right}: power spectral density of this mode. This shows a main perturbation peak around $9\times10^{-3}$\,Hz.}
    \label{fig:example_coeff}
\end{figure}

Every phase measurement in the data requires careful processing through a phase retrieval algorithm, and a certain amount of measurement noise is left in the data. Still, trends from the slew are clearly visible in the time domain. 
Looking at the power spectral density in Fig.~\ref{fig:example_coeff}, spectrum peaks between $10^{-3}$\,Hz and $10^{-2}$\,Hz are clearly visible. These correspond to the perturbations already identified\cite{Telfer2024}. The noise floor dominates above $2\times10^{-2}$\,Hz, which provides n upper floor on the real JWST data at these frequencies.

For our simulations, we do not simulate secondary mirror obstruction nor spider arms. We chose the case of EAC 3 with a primary aperture similar to JWST, with a 6.6\,m flat-to-flat aperture, but off-axis without the secondary mirror support structures. Since the dataset provides the measurements for the 18 segments from JWST, we add to the simulation a central segment that is assumed to be static.

\subsection{Data filtering and Singular Value Decomposition} \label{sec:svd}
The time series show some outliers, as visible on Fig.~\ref{fig:example_coeff} (middle), that do not seem to be realistic and are instead due to measurement artifacts. To remove these non-physical points, the 200 frames (of 11,752 total) that show the highest standard deviation are replaced by the average os the 10 surrounding data points. Then the dataset is centered by subtracting the mean over the time series of every dimension, resulting in a dataset of coefficients varying around a zero-mean value.

\begin{figure}
    \centering
    \includegraphics[width=0.8\linewidth]{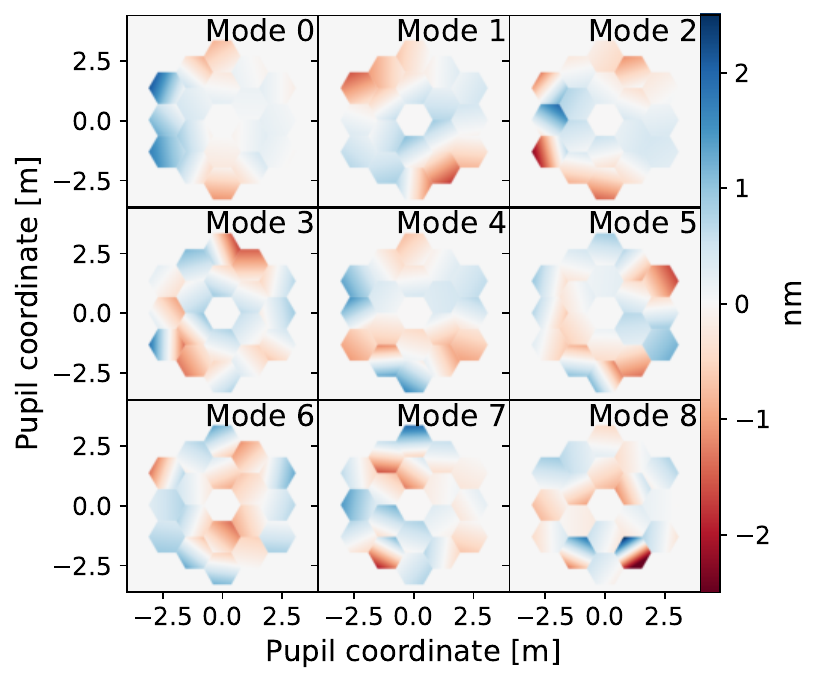}
    \caption{Surface error of the first 9 modes $\phi_i$ of the SVD of the time-series coefficients.}
    \label{fig:svd_modes}
\end{figure}

The next step is to take the SVD of the signal, yielding an eigenmode decomposition, similarly to the previous JWST analysis\cite{Telfer2024}. A view of these modes $(\phi_i)_{0\leq i<57}$ is presented on Fig.~\ref{fig:svd_modes}, with the recognizable astigmatism-shaped modes in the first components. Then the time-series piston/tip/tilt frames are projected onto the singular-mode basis. For each of the 57 PTT measurements at a given time $t$ in the time series, we get a vector of 57 coefficients $a_{i, t}$ in terms of singular modes, which constitutes our singular value time series. Since the central segment is artificially kept static, the last three modes are not meaningful. In practice, only 51 modes present a significant contribution. 

Looking at the temporal behavior of this time series, Fig.~\ref{fig:psd_world} shows the PSDs of the first 9 modes are presented. The levels are different from one mode to the other, but they all show similar features, such as high vibration modes between $10^{-3}$\,Hz and $10^{-2}$\,Hz and a similar noise floor between $10^{-5}$ and $10^{-4}$\,nm$^2$.

\subsection{Synthetic power law for PSDs} \label{sec:extrapolation}

In this section we describe how we adapt and extrapolate the power spectra to conduct our study. The PSD floor around $10^{-1}\,\mathrm{nm^2/Hz}$ visible on all modes above $10^{-2}$\,Hz on Fig.~\ref{fig:psd_world} is consistent with expected measurement noise and very likely overestimates the actual power in the PSD. In accordance to Potier et al.\cite{Potier2021}, who analyzed an observatory model and found PSDs following a power law with an exponent ranging from -1.5 to -2, we run our simulations by replacing this noise floor with a power law with an exponent of -2 above 0.02\,Hz. In addition, measurements\cite{Telfer2024} indicates other vibration peaks at higher frequencies up to 30.5\,Hz not captured by the NIRCam weak lensing telemetry producing this paper's dataset. However, knowledge of the PSD at higher frequencies is important to simulate faster control loops. Since control strategies with integral control can exhibit a resonance peak that will amplify the signal and the noise at some frequencies, it is crucial to have an appropriate estimate of the power spectra at frequencies much higher than the cutoff of the time series. As such, for a given control frequency, we extrapolate the power law with a -2 exponent up to twice the control loop frequency to not miss control-induced patterns. An example of this synthetic reconstruction of a PSD is presented on the right panel of Fig.~\ref{fig:psd_world}, with the original PSD for the mode 0 (blue), the power law approximation and extrapolation (orange dashed), the transfer function of an integral control loop (black), and the noise introduced by the control and the residual PSD after filtering (red).

\section{Instrument modeling}
Contrast predictions and control loop rejections are bound to the chosen coronagraph design and wavefront sensor architecture. The design choice for HWO is under active study\cite{Belikov2024, Belikov2026} and will depend on the exact design of the telescope, which is still an open question. In this context, we detail in this section our arbitrary design choices. The simulations were performed using the CATkit2 instrument simulation framework\cite{Catkit22026} and we rely on HCIPy\cite{Por2018} for all optical propagations. 

\subsection{Coronagraph static masks}

\begin{figure}[ht!]
    \centering
    \includegraphics[width=\linewidth]{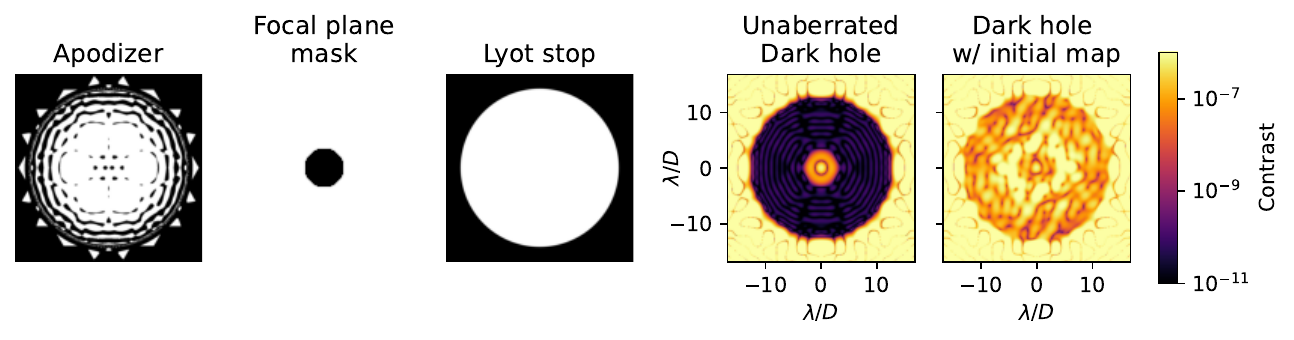}
    \caption{Coronagraphic masks of the used APLC and coronagraphic focal-plane images in the simulations. From left to right: the apodizer, from the SCDA study\cite{Nickson2022}; the focal-plane mask, with a diameter of $7\,\lambda/D$, at scale with the dark-hole images; the corresponding Lyot stop; the focal-plane intensity without aberrations, cropped to the DH region; the focal-plane contrast with the first phase screen from the time series applied on the segmented mirror.} 
    \label{fig:coronagraph_masks}
\end{figure}

The coronagraph chosen in this analysis is one of the Apodized Pupil Lyot Coronagraph (APLC) produced by the optimization tool\cite{Nickson2022} for NASA Exoplanet Exploration Program’s (ExEPs) Segmented Coronagraph Design and Analysis (SCDA) study, whose mask and example images are presented in Fig.~\ref{fig:coronagraph_masks}. It relies on an amplitude apodizer matched to the telescope primary segmented aperture, a circular focal-plane mask of diameter $7\,\lambda/D $ and a circular Lyot stop that undersizes the segmented aperture circumscribed diameter by a factor of 0.982. It is designed to produce a raw contrast better than $10^{-10}$ in a 10\,\% bandpass over the whole annular dark zone ranging from $3.4\,\lambda/D$ to $12\,\lambda/D$, with a core throughput of 25\,\% at separations further than $4.5\,\lambda/D$. 

\subsection{DMs}
For this simulation, we assume the correction is performed with a perfectly flat DM. We assume that it can correct for the segment-level aberrations that are piston, tip and tilt. Unlike the curved primary mirror, we assume it will not induce high-order aberrations coming from tilting the off-axis parabolas that are the individual hexagonal segments of the primary aperture. 
Simulating the correction with more realistic DMs, such as correcting for the segmentation with continuous face-sheet DMs, would require addressing the fitting error, since the DM cannot reproduce the sharp discontinuities of the segmentation. This will be the subject of future work. 

Nevertheless, to be able to run an Electric Field Conjugation (EFC)\cite{Giveon2007a, Giveon2007b} loop before the observing sequence, as described in Sec.~\ref{sec:efc}, it is necessary to assume that there are at least two DMs in the optical train. In our case, we simulate one in a pupil plane and the other out-of-pupil to perform phase and amplitude correction in the focal-plane electric field. For this task, we consider a pair of DMs with 48 actuators in the pupil diameter, similar in principle to those used in RST/CGI. 

\subsection{EFC solution for initialization}\label{sec:efc}
While the coronagraph design enables a contrast better than $10^{-10}$ in the dark zone with an unaberrated wavefront, the observatory will likely evolve over its lifetime, as small drifts or more abrupt events such as mechanical relaxation known as ``tilt events'' or micrometeorite impacts, described in Telfer et.al (2024)\cite{Telfer2024} will likely happen. To verify that it is possible in this situation to recover a static instrument configuration yielding the baseline contrast $E_0$, we consider that we are running EFC before starting the time-series analysis. In practice, this means that we use the two continuous DMs to recover a dark zone yielding a contrast better than $10^{-10}$ in the presence of the first aberration map of the time series. 

\begin{figure}[ht!]
    \centering
    \includegraphics[width=\linewidth]{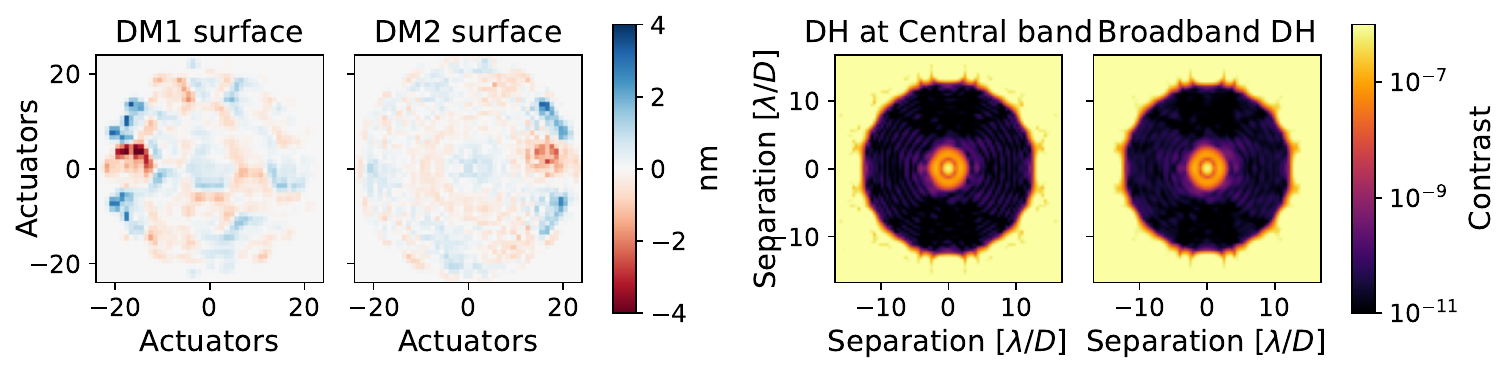}
    \caption{Example DM surface and focal-plane contrast after an EFC run of 5 iterations on the first phase screen of the time series, using the simulator-internal electric field knowledge for wavefront sensing. From left to right: DM1 and DM2 surface; focal-plane image in monochromatic light; broadband focal-plane image. The spatially averaged contrast in the central band DH is $3.5\times10^{-11}$ and the broadband one, averaged over 3 wavelength (555\,nm, 575\,nm and 595\,nm), yields $4.6\times10^{-11}$.}
    \label{fig:efc}
\end{figure}

Considering the first aberration on the time series that produces the speckle pattern visible in Fig.~\ref{fig:coronagraph_masks} (right), we show an example of the initial EFC result in Fig.~\ref{fig:efc}. With a correction of less than 10\,nm peak-to-valley on both DMs, visible on the left panels, it recovers and even improves the dark hole that is present in the two right panels. This improved contrast performance shows that it is possible to restore the $E_0$ term before starting observations.

\subsection{Wavefront sensors}
The choice of wavefront sensor has a direct impact on the performance of the control loop, as a given wavefront sensor has a given sensitivity to photon noise and read-out noise, and a given linearity range. Current state-of-the-art visible detectors achieve very low readout noise and dark current\cite{Alexani2026}, lower than 0.1 electrons. As such, to simplify our approach, we assume no readout noise in our simulations. In this work, we consider two different configurations of WFS, both using a Zernike Wavefront Sensor (ZWFS)\cite{Bloemhof2003b, Wallace2011, NDiaye2013}, known for their high sensitivity \cite{Guyon2005, Chambouleyron2021b}. One configuration uses the light rejected by the focal-plane mask, which we call the low-order WFS (LOWFS), similarly to RST/CGI\cite{Shi2016}. The other uses an unfiltered, out-of-band beam picked off by a dichroic after the apodizer plane, which we call the out-of-band WFS.

For the LOWFS case, we simulate a ZWFS with a dimple diameter of $1.06\,\lambda/D$ as previously described\cite{NDiaye2013}. Even though it lacks the high spatial frequency component of the beam due to the filtering by the focal-plane mask, it still captures some signal from the segment piston/tip/tilt aberrations, especially in a situation with a coronagraph focal-plane mask of radius of $3.5 \lambda/D$ with a pupil with 2 rings\cite{Leboulleux2026}. As demonstrated by previous work\cite{Chambouleyron2021b}, this dot diameter for the ZWFS is not optimal, but it shows an example of LOWFS usage. Optimizing for the dimple diameter would yield better sensitivity, but not beat the out-of-band configuration, which is an upper bound for the ZWFS. 

For the out-of-band WFS, we assume it is possible to pick-up light from the main optical train without introducing aberrations in the beam to the coronagraph. Even though a realistic implementation would use light at a different wavelength from the science band, we use $\lambda = 575\,$nm as well for the sake of simplicity. The exact change in photon flux will depend on the observed stellar spectrum and the observation wavelength. 

We compute an interaction matrix by poking the segmented mirror in piston, tip and tilt for each segment, with each command being normalized to produce the same wavefront error standard deviation of 10\,nm. For each mode, we perform the push and pull measurements, and every measurement is normalized by the sum of the reference measurement. The interaction matrix is then built by concatenating the differences (push minus pull) for each mode, divided by the poke amplitude for correct normalization. This interaction matrix encodes how photon noise propagates from the wavefront sensor to the modes on the segmented mirror\cite{Gendron1994a, Chambouleyron2021b, Potier2021}. 

\begin{figure}[ht!]
    \centering
    \includegraphics[width=\linewidth]{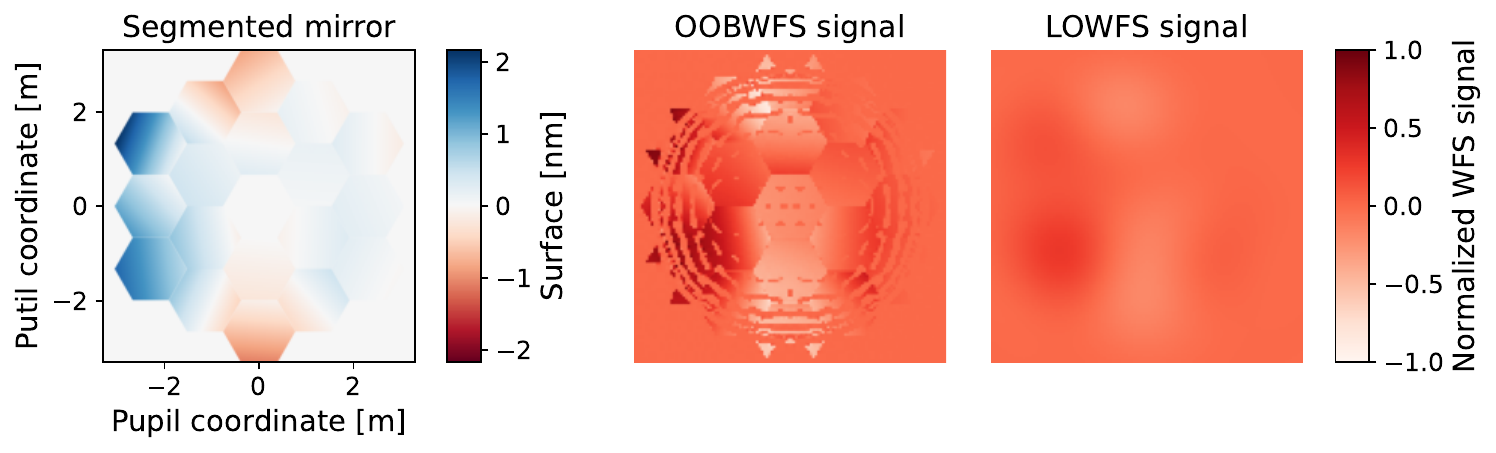}
    \caption{Examples of wavefront sensors response to the first time-series phase screen. \textbf{Left}: Surface error on the primary mirror at the first step of the time series. \textbf{Middle}: Example response of a ZWFS that has access to the whole beam before the FPM, typically an out-of-band WFS (OOBWFS). \textbf{Right}: Example response of a Low-order ZWFS using the light rejected by the Lyot FPM. The signal are represented as the difference of the measurement with the aberration minus the unaberrated measurement, normalized by the largest absolute value of the two.}
    \label{fig:wfs_responses}
\end{figure}

For the first mode of the principal component analysis, as displayed in Fig.~\ref{fig:wfs_responses} (left), we display the noiseless response of the out-of-band (middle) and the low-order (right) wavefront sensor to give a representation of the signal measured. 

\subsection{Modal contrast degradation}

\begin{figure}
     \centering
     \begin{subfigure}[b]{0.37\textwidth}
         \centering
         \includegraphics[width=\textwidth]{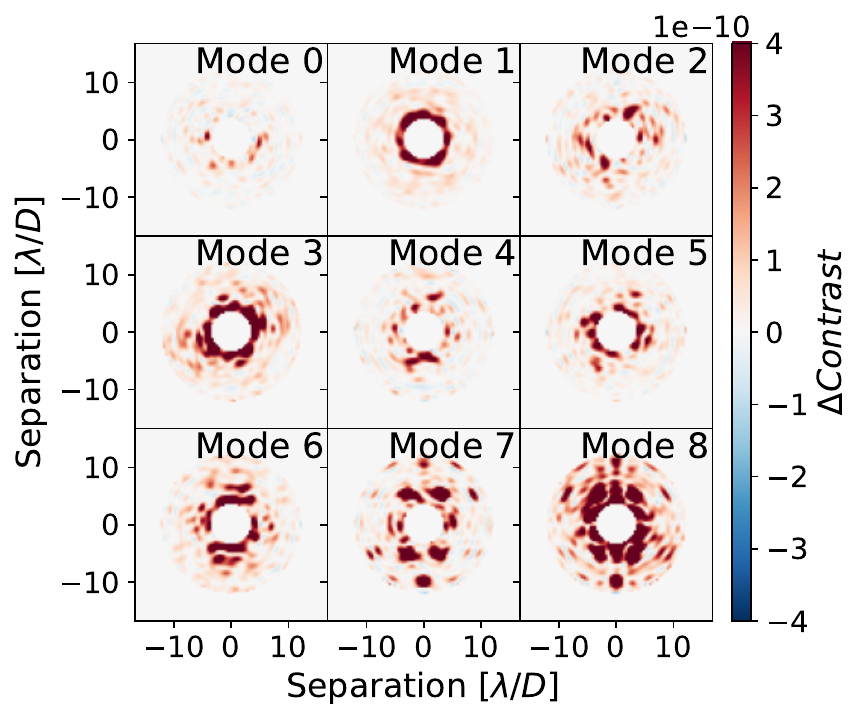}

     \end{subfigure}
     \begin{subfigure}[b]{0.62\textwidth}
         \centering
         \includegraphics[width=\textwidth]{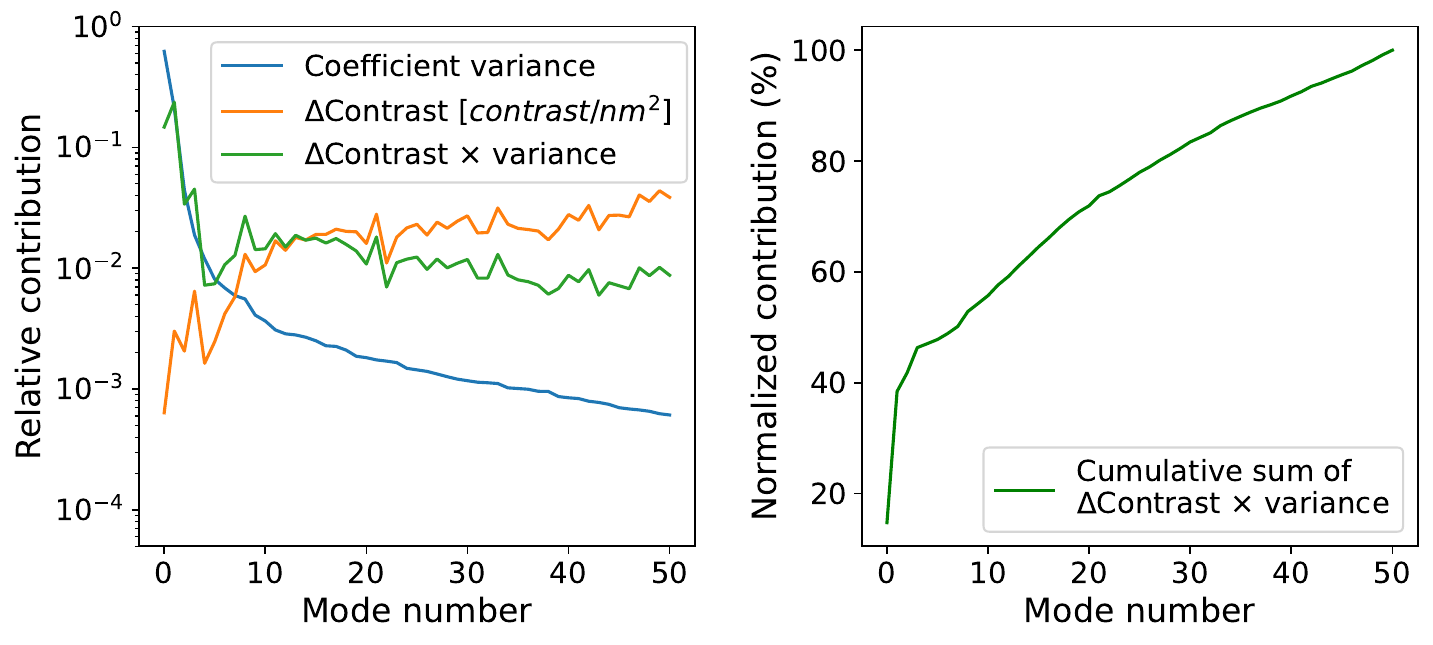}
     \end{subfigure}
        \caption{Representations of the impacts of the $(\Delta E_i)$ modes on the coronagraphic images. \textbf{Left}: Representation of the $\Delta C$ for the first 9 modes of the SVD. To isolate the exact contribution, we plot here the difference between the aberrated dark hole $|\Delta E_{seg, i} + E_0|^2$ with the unaberrated dark hole $|E_0|^2$. The wavefront aberrations are normalized to 40\,pm RMS. \textbf{Middle}: Variance (blue), delta contrast integrated over the dark zone (orange) and product of both (green) per mode, all normalized by the sum of the corresponding quantity over all the modes. \textbf{Right}: Cumulative sum of the product of the variance by the delta contrast, in percentage of the sum over all the modes.}
        \label{fig:contrast_contributions}
\end{figure}

Given this coronagraph architecture, it is possible to pre-compute the $(\Delta E_i)$ terms of Eq.~\ref{eq:residual_I}. We do so by propagating the wavefront maps from the SVD, displayed in part in Fig.~\ref{fig:svd_modes}, scaled in such a way the wavefront error is 40\,pm RMS for every mode. While displaying the $\Delta E_i$ complex maps is not straightforward, we present in linear scale in Fig.~\ref{fig:contrast_contributions} (left) the aberrated dark zones subtracted from unaberrated image $|E_0|^2$. Interestingly, for some modes, such as mode 0 or mode 2, while on average the segmented modes degrade the contrast, they can locally improve it for some speckles. Additionally, for the same wavefront error value over the pupil, the first modes, as ordered by the SVD, are not the modes that degrade the dark hole the most. This is further illustrated in Fig.~\ref{fig:contrast_contributions} (middle) where we see the variance of the modes and the $\Delta C$ curves follow opposite trends. Both combined, and as clearly visible on Fig.~\ref{fig:contrast_contributions} (right), the first five modes contribute to almost 50\,\% of the additional light in the dark hole. Then the contribution of the other modes present a fairly linear trend, meaning it is not possible to truncate the analysis to a restricted number of modes, as done for example by Potier et. al.\cite{Potier2021}.

\section{Results}

\subsection{Simulating integration over the time series}
We present in Fig.~\ref{fig:results} the contrast averaged over the dark zone in the residual contrast map $\Delta C$, computed as described in Sec.~\ref{sec:math}. It corresponds to an estimation of the residual integrated time series. In this work, we simulate two different configurations with two input perturbations. The first one is using the time series as processed in Sec.~\ref{sec:preprocessing}, that are the unscaled measurements from JWST. The second case is by dividing these coefficients by 10, done in practice by dividing the PSDs used in the first case by 100 before applying the control loop correction. While highly optimistic, this shows how a factor of 10 in stability can allow the proposed EAC3 concept to reach the desired contrast for exo-Earth imaging. 

\subsection{With unscaled JWST coefficients}

\begin{figure}
    \centering
    \includegraphics[width=\linewidth]{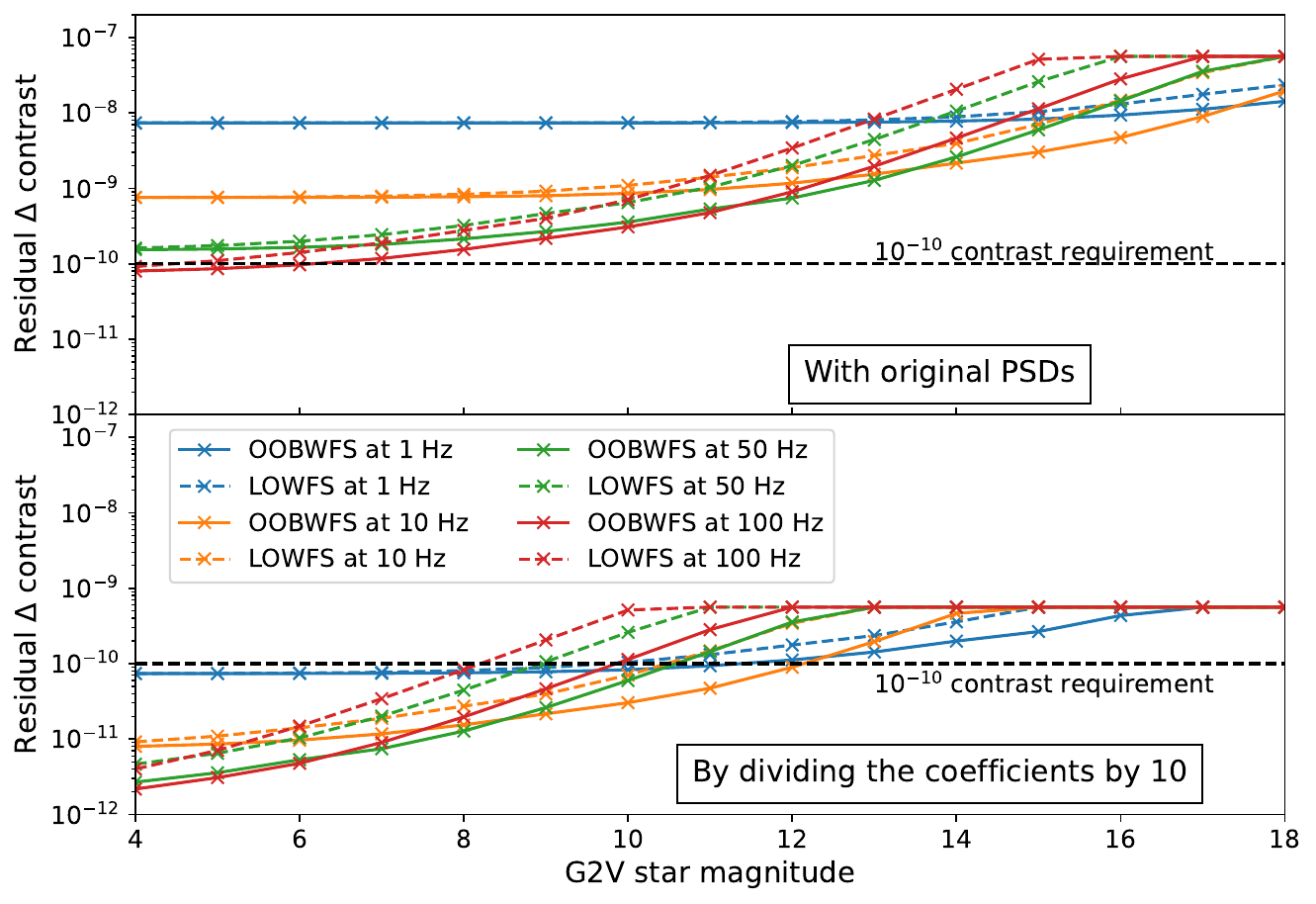}
    \caption{Simulation results presented as the spatially averaged contrast in the integrated dark zone as a function of stellar magnitude, for a LOWFS (dashed lines), and an out-of-band WFS (solid lines), running at 1\,Hz, 10\,Hz, 50\,Hz or 100\,Hz. \textbf{Top}: the PSDs used are unscaled from JWST. \textbf{Bottom}: the coefficients divided by 10.}
    \label{fig:results}
\end{figure}

In the case of the original, unscaled JWST PSDs (Fig.~\ref{fig:results}, top panel), and including a 10\,\% end-to-end throughput of the instrument, only the fastest control loop with the out-of-band WFS at 100\,Hz can reach the $10^{-10}$ contrast, and only for the stars with a magnitude of 4 or brighter. For all the loops, the plateau at the brightest stars (lowest magnitudes) means that the control is limiting and cannot be aggressive enough. Even the highest gains that do not create a resonance peak do not allow for effective corrections. For these magnitude values, the integral gain optimization usually converges to the maximum value of 0.6. Then, at higher magnitudes we are in a photon-starved regime and the control loops get dominated by the photon noise propagating to the wavefront control modes. This requires lowering the control gain, limiting the impact of the control. In these simulations, we pick 100\,Hz as the highest control frequency, but this is an arbitrary choice. The extrapolation enables probing even higher control frequencies, and going faster enables an even better rejection of the low temporal frequencies, with a transfer function as shown in Fig.~\ref{fig:psd_world} (right) shifted to the right. However, without including actual observatory data in the 10-50\,Hz regime, this makes the simulations less realistic and informative. This is further discussed in Sec.~\ref{sec:discussion}. 

\subsection{With a tenth of JWST coefficients}

In this section, we look at the hypothetical case where the coefficients of the centered time series are divided by a factor of 10, or, in other words, the PSDs are divided by 100 (Fig.~\ref{fig:results}, bottom panel). This highly optimistic case is simulated to help quantify the technological gap between JWST and HWO. In this situation, all the simulations show a delta contrast under $10^{-10}$ for star brighter than magnitude 8, showing that, under all the assumptions of this paper, this would be enough to meet the contrast needs. 
In addition, at higher magnitudes, a saturation that was not visible before is now present for all frequencies. It results from the control gain optimization which concludes that a gain of 0 is the best: the control loop introduces more variance than what it actually removes, and therefore no control is performed.

An interesting aspect, already visible for the unscaled coefficients but at higher magnitudes is the tradeoff that would have to be made between the different control speeds. While faster loop speeds perform better rejection on brighter stars thanks to the higher cutoff frequency, the higher noise floor from fainter stars makes the 10\,Hz control loop perform better between magnitudes 9 and 13. For stellar magnitudes above 13, the longest integration times perform better since they limit the amount of noise propagating in the control loop. 

\section{Discussion and limits of our results} 
\label{sec:discussion}

\subsection{Fitting error}
The work presented here relies on a number of simplifying assumptions that we now discuss. While they do not rule out the validity of our analysis they ought to be revisited in a future more granular analysis of the practical implications of the Adaptive Optics in space concept presented above. First, we neglect the fitting error introduced by the control system: that is, we assumed that the continuous DMs can correct for segment level wavefronts in real time. The correction floor for the discontinuous static wavefront is below $10^{-10}$ when the DM commands are the results of an iterative dark-hole digging algorithm, as demonstrated in Sec.~\ref{sec:efc}. However, in the continuous correction paradigm we advocate for, iterations are a major hindrance to high temporal bandwidth control. As a consequence, our results are only valid if new control algorithms are developed that operate using a modal basis that ties segment level drifts with the influence of the continuous DMs to the absolute dark hole E-field. The static response is a proof of concept that physics does not preclude the existence of such control strategies and previous work\cite{Pogorelyuk2021, Redmond2022, Redmond2024} provides a road-map. This is an exercise that we leave out for future work. In other words: applying corrections directly with the primary mirror would introduce higher-order aberrations due to its curvature, which we do not simulate here. Similarly, using continuous face-sheet mirrors would not correct for the highest spatial frequencies introduced by segment-level errors. In the first case, a possible mitigation would be to split the correction between two deformable mirrors: a flat segmented one to address segmentation effects, and a continuous one for higher-order, segment-level modes. In the second case, using deformable mirrors with a sufficient number of actuators, such as RST's 48$\times$48 DMs, would in principle enable full control within the spatial frequencies relevant to the dark zone, leaving fitting errors outside of it. However, higher-order nonlinear effects and coupling between modes could play a role and results in this paper are only valid under the assumption of the implementation of control strategies that mitigate these terms.

\subsection{Vibrations}
A second consideration concerns the sampling frequency of the time series. Simulating control loops that run slower than a sampling frequency of 0.6\,Hz provides little additional insight, since the effective bandpass of the filter then allows only marginal correction: an attenuation of the PSD by a factor of 100, equivalent to a factor of 10 in the coefficients, is only reached at frequencies tens of times below the control frequency, as shown by the transfer function in the right panel of Fig.~\ref{fig:psd_world}. For this reason, we run our simulations with loop frequencies well above the sampling frequency of the time series. Extrapolating the PSD beyond the sampling rate is nonetheless necessary. In Sec.~\ref{sec:extrapolation}, we simply extrapolated the PSD slope in the region dominated by JWST sensing noise. This leaves our analysis blind to any real mechanical vibrations or resonances, and their potential amplification introduced by the control loop. Examples of such open-loop vibrations are common and can be found in ground-based adaptive optics systems\cite{Poyneer2016}, finite-element simulations of the LUVOIR concept \cite{Carrier2025, Potier2021}, and in the JWST data \cite{Telfer2024}. In ground-based observatories, they often arise from power sources or observatory cooling sub-systems (e.g., fans). In space-based telescopes, they can arise, for instance, from reaction wheels, and occur on JWST at frequencies between 10\,Hz and 35\,Hz. Our integrator-based control loop, which we limit to 100\,Hz, cannot efficiently reject these vibrations. Adding such vibration peaks to the PSDs in Fig.~\ref{fig:psd_world} by spreading them across all 51 modes considered, with a total power of 3\,nm, degrades the achievable contrast by a factor of more than a hundred, even on bright stars with a control loop running at 2000\,Hz, orders of magnitude above the state-of-the-art loop stabilization frequency on JWST\cite{Lallo2022}.

In practice, however, it is unlikely that the vibration would couple into all of the primary mirror modes, and running a simple integrator is not the appropriate solution for well-defined, peaked disturbances. It has been demonstrated from the ground, on sky\cite{Poyneer2016}, that a specific Kalman filter can reduce vibration peaks by a factor of 10 in amplitude without degrading the rest of the PSD. Applied to the vibrations measured on JWST, this approach could therefore reduce the total vibration to only 300\,pm. In addition, by keeping the 3\,nm total vibration but assuming that the time series coefficients have been reduced by a factor of 30 (corresponding to a PSD attenuation of 900), our integrator can still reach a contrast of $10^{-10}$ when running at 2000\,Hz on stars brighter than magnitude 4. This therefore suggests a dual pathway: not only can more advanced control strategies, such as Kalman filters, specifically mitigate vibrations, but structural improvements to observatory stability could also help address them. A more precise estimate would require further trade-off analysis, which we leave for future work.

\section{Conclusion}

This contribution investigates the potential achievable raw contrast from a JWST-like telescope if it was equipped with active wavefront control and a modern coronagraph design, to quantify the technology gap left towards Earth-like planet imaging around Sun-like stars with HWO. We combined in-flight measurements of the JWST primary-mirror wavefront, acquired during commissioning, with an end-to-end numerical model of a segmented, unobstructed telescope equipped with an apodized pupil Lyot coronagraph, deformable mirrors, and photon-noise-limited Zernike wavefront sensors. By computing the eigenmodes of the measured segment piston, tip, and tilt coefficients and propagating their temporal power spectral densities through a closed-loop integrator model, we estimated the residual stellar leakage accumulated over science observations lasting several hours.

At the measured JWST disturbance levels, active control substantially reduces the additional stellar leakage but does not consistently maintain a spatially averaged raw contrast below $10^{-10}$, particularly for stars fainter than magnitude 6. However, reducing the amplitudes of the measured segment motions by a factor of ten, corresponding to a factor of 100 reduction in their power spectral densities, enables both the out-of-band and low-order wavefront-sensing configurations to maintain a raw contrast below $10^{-10}$ for Sun-like G2V stars brighter than magnitude 11, under the assumptions adopted here. This result, based on in-flight data, provides a new estimate of the stability gap that must be bridged between JWST and HWO. In particular, it illustrates how active control can relax the structural stability requirements imposed on the observatory.

These conclusions depend strongly on the temporal-frequency content of the disturbances. In particular, extrapolating the measured low-frequency spectra does not capture the reaction-wheel disturbances observed on JWST between approximately 10\,Hz and 35\,Hz. A conservative simulation in which these vibrations are coupled equally into all controlled modes degrades the predicted contrast by approximately two orders of magnitude. Under these conditions, an integral controller would require kilohertz-class operation and very bright targets to approach a raw contrast of $10^{-10}$. This pessimistic case highlights both the importance of limiting vibration coupling through the observatory design and the need for predictive controllers, such as linear-quadratic-Gaussian controllers, targeted at narrow-band disturbances.

The present analysis remains an idealized estimate. It assumes an unobstructed off-axis architecture, perfect correction of segment piston, tip, and tilt, negligible detector noise, and simplified models of out-of-band wavefront sensing. Nevertheless, the framework developed here directly connects measured observatory telemetry, wavefront-sensor sensitivity, controller architecture, and coronagraphic performance. Moreover, it considers the spatially averaged raw contrast which is a conservative metric. A dark hole maintained at a raw contrast of $10^{-10}$ over an observing sequence can enable detections at deeper effective contrasts after post-processing, particularly when diversity is provided by telescope roll maneuvers during observations\cite{Soummer2012}. The gain to be expected depends on the spatio-temporal correlation structure of the residual speckles: a reproducible residual halo is easier to remove than a more rapidly varying pattern. As such, an exact gain value for this study is not straightforward to obtain, but recent studies have demonstrated signal-to-noise improvements by a factor of more than eight\cite{Redmond2024}, for example. While further work is needed to quantify the achievable gain with our approach, we expect the potential improvements to be significant. Despite these approximations, this study constitutes one of the first performance estimates based on measurements from an existing observatory rather than exclusively on observatory models.

Overall, our results indicate that JWST’s stability provides an encouraging demonstration in the context of HWO development, although JWST-level segment stability combined with active wavefront control is likely insufficient to guarantee a raw contrast of $10^{-10}$. Under the assumptions considered here, an approximately tenfold reduction in wavefront amplitudes, combined with active control techniques inherited from ground-based adaptive optics and dedicated mitigation of narrow-band vibrations, can substantially narrow this gap. Achieving this improvement appears plausible because the sources of the dominant disturbances, including reaction wheels and electronic heating systems, have already been identified, and existing mitigation strategies could be implemented in the design of a future observatory. Future work will incorporate more realistic estimates of the contributions from vibrations and detector readout noise, as well as targeted modal control and more realistic deformable-mirror models. Combined with estimates of post-processed sensitivity, these developments will help reduce the uncertainty associated with the remaining technological gap that must be bridged to detect exo-Earth signals.

\subsection*{Disclosures}
The authors declare that there are no financial interests, commercial affiliations, or other potential conflicts of interest that could have influenced the objectivity of this research or the writing of this paper

\subsection*{Code, Data, and Materials Availability} 
The timeseries data are from the observation program 1445 from JWST, publicly available on the Mikulski Archive for Space Telescopes (MAST). The simulation made use of the HCIPy\cite{Por2018, Por2026hicpy2}, NumPy\cite{Numpy}, Matplotlib\cite{Matplotlib}, AstroPy\cite{Astropy3, Astropy2, astropy1}, SciPy\cite{SciPy}, Catkit2\cite{Por2026catkit2} and Exoscene packages. 

\subsection*{Acknowledgments}
R. P. was partly supported by Grant 80NSSC22K0372 issued through the Astrophysics Research and Analysis Program (APRA; PI: L. Pueyo). E.H.P. was supported in part by the NASA Hubble Fellowship grant HST-HF2-51467.001-A awarded by the Space Telescope Science Institute, which is operated by the Association of Universities for Research in Astronomy, Incorporated, under NASA contract NAS5-26555. E.H.P. acknowledges support from the Heising-Simons Foundation through the 51 Pegasi b Fellowship. S.S. acknowledges support by STScI Postdoctoral Fellowship. The authors acknowledge the use of Anthropic's Claude and ChatGPT Sol for sentence formulation, spelling and grammar check. The content and calculations did not involve AI.


\bibliography{report} 
\bibliographystyle{spiejour}  


\vspace{2ex}\noindent\textbf{Raphaël Pourcelot} is a postdoc at MPIA, Germany within the Planet Formation and Exoplanet department. This succeeds his first postdoc at STScI, within the Russell B. Makidon Optics Laboratory. He received his PhD from the Université Côte d’Azur in Nice, France in 2022. He is mainly working on wavefront sensing and control for high-contrast imaging for exoplanet detection and Zernike wavefront sensor applications in particular.

\vspace{2ex}\noindent\textbf{Laurent A. Pueyo} is an associate astronomer at Space Telescope Science Institute (STScI). He received his doctorate from Princeton University in 2008; he previously worked as a NASA fellow at JPL, California, and was a Sagan fellow at Johns Hopkins University (JHU). His research focuses on imaging faint planets around nearby stars. He has pioneered optical technologies that allow astronomers to take images of other planetary systems and has developed data analysis methods now standardly used to study extrasolar planets.

\vspace{2ex}\noindent\textbf{Emiel H. Por} is an unaffiliated researcher. He was a 51 Pegasi B fellow at the University of California Santa Cruz (UCSC). Before that, he was an NASA Sagan fellow at the Space Telescope Science Institute (STScI) in Baltimore. He received his doctorate in astronomy from Leiden University, the Netherlands. His research interests include coronagraphy, integrated photonics, wavefront sensing and control and high-contrast imaging, with a particular interest in simulation and control software.

\vspace{2ex}\noindent\textbf{Sarah Steiger} is an Astronomical Optics Scientist at STScI working with the Telescopes Branch and Russell B. Makidon Optics Laboratory. She received her doctorate from the University of California, Santa Barbara in 2023 and is an expert in high-contrast imaging and superconducting detector technologies. Her current focus is on JWST wavefront sensing and the development of new technologies to enable the exoplanet imaging goals of future flagship space observatories, including detectors, coronagraphs, and wavefront sensing and control systems.

\vspace{2ex}\noindent\textbf{Rémi Soummer} is an astronomer at STScI. He received his doctorate from the University of Nice in 2002 and has been working in the field of high-contrast imaging and instrumentation for the detection and characterization of exoplanets ever since. He is currently the head of the Russell B. Makidon Optics Laboratory and working on a coronagraph demonstration for future large segmented aperture space telescopes.

\vspace{2ex}\noindent\textbf{Iva Laginja} is an associate researcher in astronomical instrumentation with CNRS at the Laboratoire Lagrange in Nice, France. Previously, she was an astronomical optics scientist at the Space Telescope Science Institute and a postdoctoral researcher at LESIA in Paris. She received her doctorate degree in astronomy and instrumentation at STScI/ONERA/LAM.

\vspace{1ex}
\noindent Biographies of the other authors are not available.

\listoffigures
\listoftables

\end{document}